\documentclass[a4paper,10pt]{article}
\usepackage[utf8x]{inputenc}
\usepackage{graphicx, color}
\usepackage{amsmath}
\usepackage{amscd}
\usepackage{titlesec}
\usepackage{listings}
\usepackage{enumerate}
\usepackage{amssymb}
\usepackage{amsthm}
\usepackage{syntonly}
\usepackage{fancyhdr}
\usepackage{hyperref}
\newcommand{\ket}[1]{\left|#1\right>}

\title{Calculation of the determinant in the Wheeler-De Witt equation.}
\author{Carlos A. Jim\'enez-Orjuela\\
Nelson Vanegas-Arbel\'aez\\
Instituto de F\'isica\\
Universidad de Antioquia.
\vspace{1cm}\\
\textit{cajimen@pegasus.udea.edu.co}\\
\textit{nvanegas@fisica.udea.edu.co}}

\begin{document}

\maketitle

\begin{abstract}
The Riemann-zeta function regularization procedure has been studied intensively as a good method in the computation of the determinant for
pseudo-diferential operator. In this paper we propose a different approach for the computation of the determinant base on the 
Wheeler-De Witt equation.
\end{abstract}

\section{Introduction}

In this theory, the space has geometry $\mathbb{R}\times\Sigma$ where $\Sigma$ is a two dimensional compact manifold, usually a 2-Torus $T^2$.
The eignestates and eigenvalues are \cite{scarlip}:
\begin{equation}
 \ket{m\,n}\,=\, e^{2\pi i(mx+ny)}\hspace{1cm}l_{mn}=\frac{4\pi^2}{\tau_2}|n-m\tau|^2+V_0,
\end{equation}

\noindent where $m,n$ are integers, and $x,y$ angular coordinates, $V_0$ is related to the potential function and $\tau=\tau_1+i\tau_2$ is the
complex modulus asociated to a metric \cite{elizalde},

\begin{equation}
d\bar{s}^2=\frac{1}{\tau_2}|dx+\tau dy|^2.
\end{equation}

The zeta function asociated with the operator is;

\begin{equation}
\zeta(s)=\sum_{n,m}(l_{nm})^{-s}.
\end{equation}

\noindent Then, the determinant of the operator is:
\begin{eqnarray}
\mbox{det}\,\,D_0 & = & \prod_{n,m}l_{nm} = \prod_{n,m}e^{ln(l_{nm})}\nonumber\\
   & = & \mbox{exp}\left[-\lim_{s\rightarrow 0}\frac{d}{ds}\sum_{nm}l_{nm}^{-s}\right]\nonumber\\
   & = & e^{-\zeta'(0)}.
\end{eqnarray}

\noindent In \cite{nvanegas} we find that the zeta function is asociated to the Gamma function as:
\begin{equation}
\zeta(s)=\sum_{n}l_{nm}^{-s}=\frac{1}{\Gamma(s)}\int_{0}^{\infty}t^{s-1}\sum_{nm}e^{-l_{nm}t}dt,
\end{equation}
with,

\begin{equation}
 e^{-l_{nm}t}=e^{V_0t}e^{-\frac{4\pi^2}{\tau_2}|n-m\tau|^2t},
\end{equation}
and,
\begin{eqnarray}
 |n-m\tau|^2&=&\left(n-m(\tau_1-i\tau_2)\right)\left(n-m(\tau_1+i\tau_2)\right)\nonumber\\
   &=& n^2-2nm\tau_1+m^2(\tau_1^2+\tau_2^2).
\end{eqnarray}
This can be wrote in a matrix form as:

\begin{eqnarray}
  |n-m\tau|^2 & = & \left( n\,\, , \,\, m\right) \cdot
  \left(
  \begin{array}{cc}
    1  &   -\tau_1  \\ -\tau_1 & \tau_1^2+\tau_2^2
  \end{array}  \right) \cdot
  \left(
       \begin{array}{c} n\\m \end{array} \right), \nonumber \\[8pt]
  & \equiv & \vec{n}^t \,\, \Omega' \,\, \vec{n}  \label{eq:cuad},
\end{eqnarray}
this matrix satisfies:

\begin{equation}
 \det \, \Omega' \, =  \, \tau_2^2.
\end{equation}
Then we can define a new matrix as $\Omega=\Omega'/\tau_2$, where $\det \, \Omega \, = 1$, and the zeta function is:

\begin{eqnarray}
\zeta(s)&=&\frac{1}{\Gamma(s)}\int_{0}^{\infty}dt\,t^{s-1}\sum_{\vec{n}}e^{-4\pi^2\vec{n}^T\Omega\vec{n}t-V_0t}\nonumber\\
   &=& \frac{1}{\Gamma(s)}\int_{0}^{\infty}dt\,t^{s-1}e^{-V_0t}\sum_{nm}e^{\frac{-4\pi^2t}{\tau_2}A_{nm}}\label{eq:mizeta},
\end{eqnarray}
with,

\begin{eqnarray}
 A_{nm}&\equiv& n^2-2nm\tau_1+m^2(\tau_1^2+\tau_2^2)\nonumber\\
   &=&  \left(n^2-m\tau_1\right)^2+m^2\tau_2^2.
\end{eqnarray}

\section{Some Useful Results}

Let's list three equation we are going to use \cite{nvanegas, chan, table} 

\begin{eqnarray}
 \Psi\equiv\lim_{s\rightarrow0}\frac{d}{ds}\zeta(s)&=&\lim{s\rightarrow0}\left(\Gamma(s)\zeta(s)+\frac{1}{s}\right),\nonumber\\
 \sum_{n\in\mathbb{Z}}e^{-\pi t(n+v)^2}&=&\frac{1}{\sqrt{t}}\sum_{n\in\mathbb{Z}}e^{2\pi inv-\frac{\pi n^2}{t}}\hspace{1cm} t>0,\nonumber\\
 \int_0^\infty dx \,x^{\nu-1}\, e^{-\frac{\beta}{x} -\alpha x } &=& 2\,\Big(\frac{\beta}{\alpha}\Big)^{\nu/2}\,  K_{\nu}(2\sqrt{\beta\alpha}),\label{eq:list}
\end{eqnarray}
for the computation ahead, with this in mind and following in the same line as in \cite{nvanegas},  we now split the problem in four contributions, 

\begin{equation}
\zeta(s) = \hspace{0.3cm}\underbrace{\zeta_1(s)}_{n=m=0}\hspace{0.3cm}+\hspace{0.3cm}\underbrace{\zeta_2(s)}_{n=0,m\neq0}
\hspace{0.3cm}+\hspace{0.3cm}\underbrace{\zeta_3(s)}_{m=0,n\neq0}+\hspace{0.3cm}\underbrace{\zeta_4(s)}_{m\neq0,n\neq0}.\nonumber\\
\end{equation}

\noindent The first case is solved making $(n=m=0)$ in the integral (\ref{eq:mizeta}), then we can see that under $x=V_0t$, 
the gamma function is canceled, 

\begin{eqnarray}
 \zeta_1(s)&=&\frac{1}{\Gamma(s)}\int_{0}^{\infty}dt\,t^{s-1}e^{-V_0t}\nonumber\\
   & = & V_0^{-s},
\end{eqnarray}
then,
\begin{equation}
 \lim_{s\rightarrow0}\zeta_1 '(s)=-\mbox{ln}V_0.
\label{eq:primero}
\end{equation}

\noindent The second case, with ($n=0$) in (\ref{eq:list}) is:

\begin{equation}
\zeta_2(s) = \frac{1}{\Gamma(s)}\int_{0}^{\infty}dt\,t^{s-1}e^{-V_0t}
\sum_{m\neq0}e^{-\alpha\pi m^2t}\hspace{1cm}\alpha\equiv\frac{4\pi\tau^2}{\tau_2}.
\end{equation}
If we consider $\tau_2>0$, and using (\ref{eq:list}), we obtain,

\begin{eqnarray}
\zeta_2(s) &=& \frac{1}{\Gamma(s)}\sum_{m\neq0}\int_{0}^{\infty}dt\,\frac{t^{s-1}}{\sqrt{\alpha t}}e^{-V_0t}
e^{-\frac{\pi m^2}{\alpha t}}\nonumber\\
   & = & \frac{1}{\Gamma(s)\sqrt{\alpha}}\sum_{m\neq0}\int_{0}^{\infty}dt\,t^{(s-1/2)-1}e^{-\frac{\pi m^2}{\alpha t}-V_0t}.
\end{eqnarray}

\noindent And with (\ref{eq:list}), if $V_0>0$ then,

\begin{equation}
 \zeta_2(s) = \frac{2}{\Gamma(s)\sqrt{\alpha}}\sum_{m\neq0}\left(\frac{\pi m^2}{V_0\alpha}\right)^{(s/2-1/4)}
K_{s-1/2}\left(2\sqrt{\frac{\pi m^2V_0}{\alpha}}\right).\label{eq:segundo}
\end{equation}

\noindent The third contribution, with ($m=0$) in (\ref{eq:mizeta}) is:

\begin{equation}
\zeta_3(s) = \frac{1}{\Gamma(s)}\int_{0}^{\infty}dt\,t^{s-1}e^{-V_0t}
\sum_{n\neq0}e^{-\gamma\pi n^2t}\hspace{1cm}\gamma\equiv\frac{4\pi}{\tau_2}.
\end{equation}
And again we consider $\tau_2>0$ for use (\ref{eq:list}) to obtain:

\begin{eqnarray}
\zeta_3(s) & = & \frac{1}{\Gamma(s)\sqrt{\gamma}}\sum_{n\neq0}\int_{0}^{\infty}dt\,t^{(s-1/2)-1}e^{-\frac{\pi n^2}{\gamma t}-V_0t}\nonumber\\
  & = & \frac{2}{\Gamma(s)\sqrt{\gamma}}\sum_{n\neq0}\left(\frac{\pi n^2}{V_0\gamma}\right)^{(s/2-1/4)}
K_{s-1/2}\left(2\sqrt{\frac{\pi n^2V_0}{\gamma}}\right).\label{eq:tercero}
\end{eqnarray}

\noindent The fourth contribution, with ($m,n\neq0$) in (\ref{eq:mizeta}) and again using (\ref{eq:list}) we obtain:

\begin{equation}
\zeta_4(s)=\frac{\sqrt{\tau_2}}{\sqrt{\pi}\Gamma(s)}\sum_{n,m\neq0}\left(\frac{\tau_2 n^2}{4V_0+16\pi^2 m^2\tau_2}\right)^{s/2-1/4}
K_{s-1/2}\left(\sqrt{\frac{\pi^2\tau_2 n^2(V_0+4\pi^2m^2\tau_2)}{\pi^2}}\right).\label{eq:cuarto}
\end{equation}

\section{Final Results and Conclusions}

Putting together the contributions found in the last secction, replacing the parameters ($\alpha$, $\gamma$)
and taking the limit in (\ref{eq:list}), we obtain the expression

\begin{eqnarray}
 \Psi & = & \sqrt{\frac{\tau_2}{\pi}}\frac{1}{|\tau|}\sum_{n\neq0}\left(\frac{4|\tau|^2V_0}{n^2\tau_2}\right)^{1/4}
K_{-1/2}\left(\sqrt{\tau_2V_0}\left|\frac{n}{\tau}\right|\right)\nonumber\\
   & + & \sqrt{\frac{\tau_2}{\pi}}\sum_{n\neq0}\left(\frac{4|V_0}{n^2\tau_2}\right)^{1/4}
K_{-1/2}\left(\sqrt{\tau_2V_0}|n|\right)\nonumber\\
& + & \sqrt{\frac{\tau_2}{\pi}}\sum_{n,m\neq0}\left(\frac{4(V_0+4\pi^2m^2\tau_2)}{n^2\tau_2}\right)^{1/4}
K_{-1/2}\left(\sqrt{\tau_2V_0+4\pi^2m^2\tau_{2}^{2}}|n|\right)\nonumber\\
&-&\mbox{ln}V_0.
\end{eqnarray}
The integers ``n,m`` are non zero, then we can write 
\begin{equation}
\sum_{n\neq0}f(|n|)=2\sum_{n=1}^{\infty}f(n).\hspace{1cm}\sum_{n,m\neq0}f(|n|,|m|)=4\sum_{n,m=1}^{\infty}f(n,m).
\end{equation}
after some work and using some Bessel function identities we obtain

\begin{eqnarray}
 \Psi & = & 2|\tau_2|^{1/4}\sqrt{\frac{2}{\pi}}\sum_{n=1}^{\infty}\bigg[\frac{\sqrt{|\tau|}V_{0}^{1/4}}{\sqrt{n}}
K_{1/2}\left(\frac{n\sqrt{V_0\tau_2}}{|\tau|}\right)+\frac{V_{0}^{1/4}}{\sqrt{n}}K_{1/2}\left(n\sqrt{V_0\tau_2}\right)\nonumber\\
 & + & 2\sum_{m=1}^{\infty}\frac{(V_{0}+4\pi^2m^2\tau_2)^{1/4}}{\sqrt{n}}K_{1/2}\left(n\sqrt{V_0\tau_2+4\pi^2m^2\tau_{2}^{2}}\right)
\bigg]-\mbox{ln}V_0.
\end{eqnarray}

Then the determinant we are looking for is:
\begin{eqnarray}
 \det\,D_{0}&=&V_0 \exp \Bigg\{-2|\tau_2|^{1/4}\sqrt{\frac{2}{\pi}}\sum_{n=1}^{\infty}\bigg[\frac{\sqrt{|\tau|}V_{0}^{1/4}}{\sqrt{n}}
K_{1/2}\left(\frac{n\sqrt{V_0\tau_2}}{|\tau|}\right)\nonumber\\
&+& \frac{V_{0}^{1/4}}{\sqrt{n}}K_{1/2}\left(n\sqrt{V_0\tau_2}\right)\nonumber\\
&+& 2\sum_{m=1}^{\infty}\frac{(V_{0}+4\pi^2m^2\tau_2)^{1/4}}{\sqrt{n}}K_{1/2}\left(n\sqrt{V_0\tau_2+4\pi^2m^2\tau_{2}^{2}}\right)\bigg]
\Bigg\}.\label{eq:finalwdw}
\end{eqnarray}

\noindent The last expression is regular, because the Bessel function is asimptotic and the expression is multiplied by a factor 
$1/\sqrt{n}$, then the important terms are the first in the series, at this point we will compute that limit with the
help of the identity \cite{arfken}:

\begin{equation}
\mbox{lim}_{z\rightarrow0}K_{\nu}(z)=\frac{(\nu-1)!2^{\nu-1}}{z^{\nu}}\hspace{1cm}\nu>0.
\label{eq:id-asintoticas}
\end{equation}

For the first values of $m$, and $n\ll1/\sqrt{V_0\tau_2}$ we obtain:

\begin{equation}
 \Psi_{z\ll1}\approx \frac{2|\tau|+6}{n}-\mbox{ln}V_0,
 \end{equation}
It does\'nt mean that the series converg to this value, but than this series can be wrote like a sum of the first terms of this kind.\\ 
In this case the determinant is,
\begin{equation}
 \mbox{det}D_0\approx V_0 \sum_{\mbox{first n's}}\mbox{exp}\left[\frac{-2|\tau|-6}{n}\right].
\end{equation}

Our final result (\ref{eq:finalwdw}) is diferent that some results found in the literature, 
for example in \cite{elizalde, reg-elizalde, some}. Maybe there is some limit in wich these result are equivalent, or maybe really there 
is diferent in a deep sense, the important thing is than it\'s limit seems more simple and handle and its well bahaviour is more evident.
This diference may have some interesting consecuences on some works derived from the quantization with the Wheeler-De Witt equation, in particular
in brane worlds models, like in \cite{brane1}, where the Wheeler-De Witt equation is solved with first-class Dirac constraints defined in a five 
dimensional space, and the important determinant here is known as Morette-DeWitt determinant, wich must be solved. other nice work is \cite{brane2}, where
is considered the gravity on the worldvolume
of D3-brane embedded in the flat background produced
by N p-branes, first clasically, and later in the quantization, the Wheeler-De Witt equation must be solved.

\end{document}